\documentclass[aps,pra,amsmath,twoside, showpacs,reprint,floatfix]{revtex4-2}

\usepackage{xspace}
\usepackage{graphicx}
\usepackage{dcolumn}
\usepackage{braket}

\newcommand{\rmat}{\textit{R}-matrix\xspace}
\newcommand{\rsmat}{\textit{R} matrix\xspace}

\newcommand{\bspls}{\textit{B}-splines\xspace}

\newcommand{\figab}{Fig.~}
\newcommand{\figsab}{Figs.~}
\newcommand{\eqab}{Eq.~}

\newcommand{\eqsab}{Eqs.~}

\newcommand{\secab}{Section~}
\newcommand{\electron}{e$^-$\xspace}

\newcommand{\tsigma}{$^3\Sigma$\xspace}
\newcommand{\rbtwo}{Rb$_2$\xspace}
\newcommand{\rbtwoplus}{Rb$_2^+$\xspace}
\newcommand{\rbminus}{Rb$^-$\xspace}
\newcommand{\rbplus}{Rb$^+$\xspace}
\newcommand{\cora}{$A$\xspace}
\newcommand{\corb}{$B$\xspace}
\newcommand{\schr}{Schr\"odinger\xspace}

\begin{document}
	\title{Long-range Rydberg molecule Rb$_2$: Two-electron \rmat calculations at intermediate internuclear distances}
	\author{\firstname{Michal} Tarana}
	\affiliation{J. Heyrovsk\'y Institute of Physical Chemistry of the Czech Academy of Sciences, Dolej\v{s}kova 
	2155/3, 182 23 
	Prague 8, Czech Republic}
	\email{michal.tarana@jh-inst.cas.cz}
	\begin{abstract}
		The adiabatic potential energy curves of \rbtwo in the long-range Rydberg electronic states are calculated 
		using the two-electron \rmat method [M. Tarana, R. \v{C}ur\'{i}k, Phys. Rev. A \textbf{93}, 012515 (2016)] for 
		the intermediate internuclear separations between 37 
		a.u. and 200 a.u. The results are compared with the zero-range models to find a region of the internuclear 
		distances where the Fermi's pseudopotential approach provides accurate energies. A finite-range potential
		model of the atomic perturber is used to calculate the wave functions of the Rydberg electron and their 
		features specific for the studied range of internuclear distances are identified.
	\end{abstract}
	\maketitle
	\section{Introduction}
		Diatomic long-range Rydberg molecule (LRRM) is an exotic system of two atoms -- one in its ground state and 
		one in its high excited state (typically $n\sim 10 - 80$), bound to each other at the distances of the nuclei 
		varying 
		between tens and 
		thousands of Bohr radii. The mechanism of this bond consists in the scattering of the Rydberg electron off 
		the distant neutral atom (perturber). This interaction can affect the phase of the Rydberg wave function in 
		such way that the molecular electronic bound state is formed. When its energy, as a function of the 
		internuclear distance, forms sufficiently deep well, the vibrational states of the LRRMs can be bound.
		
		Existence of the LRRMs was first theoretically predicted by \citet{Greene-prl} almost two decades ago along 
		with their unusually large permanent electric dipole moments, even for the homonuclear diatomic molecules. Two 
		categories of the electronic bound states were identified: Those formed by the perturbation of the 
		non-degenerate atomic Rydberg states with low angular momenta and the trilobite states involving the 
		hydrogen-like degenerate atomic states with high angular momenta. It took almost nine years since then until 
		the first experimental evidence of the LRRMs was provided by \citet{Bendkowsky2009} and their electric dipole 
		moment measured by \citet{Li2011-sci}.
		
		Since then, the LRRMs became a subject of intensive theoretical and experimental research. The trilobite-like 
		states were observed in cesium by \citet{Booth2015sci}. The existence of the butterfly states, predominated by 
		the $p$-wave interaction of the Rydberg electron with the neutral perturber, predicted by \citet{Hamilton2002}, 
		was experimentally confirmed by \citet{Niederprum2016nat}. The LRRMs have been so far predominately 
		prepared in the ultracold atomic ensembles of the heavy alkali metals 
		\cite{Bendkowsky2009,Booth2015sci,Niederprum2016nat}. However, \citet{DeSalvo2015}, and more recently 
		\citet{Ding2019}, reported successful creation of the LRRMs in the ultra-cold strontium gas.
		
		The LRRMs have also provided ways to explore other phenomena. \citet{Schmid2018} proposed an experiment 
		where the ionized LRRM 	Li$_2$ provides a well defined initial state for the Li-Li$^+$ collision in the 
		quantum regime that was not available to previously developed experimental techniques. The LRRMs \rbtwo have 
		also been utilized to study the effects of the spin-orbit interactions in the electron collisions with 
		the rubidium atom at low scattering energies \cite{Engel2019,Deiss2020}.
		
		Summary of the related research exceeds the scope of this article. For comprehensive review, see the recent 
		papers \cite{Fey2020rev,Shaffer2018,Eiles2019} and references therein.
		
		The first and so far most frequently utilized theoretical model of the LRRMs is based on the representation of 
		the 
		neutral perturber by the Fermi's zero-range pseudo-potential \cite{Greene-prl,Hamilton2002} that couples the 
		atomic 
		eigenstates 
		of the Rydberg electron. This delta-function interaction possesses a singularity at the 
		position of the perturber. It is usually considered non-zero only in the partial waves $s$ and $p$ and 
		parameterized by the 
		generalized energy-dependent electron-perturber scattering length \cite{Greene-prl,blatt-inbook} in the 
		$s$-wave. Following \citet{Omont1977}, the $p$-wave component, particularly important for the alkali metals 
		supporting the low-energy $^3P^o$ resonance, is parameterized by the low-energy $p$-wave phase shift of the 
		electron-perturber scattering \cite{Hamilton2002}. This simple model was more recently enhanced by taking the 
		spin effects into account \cite{Anderson,Eiles2017} and provided an insight into several experimental results. 
				
		The computational method beyond the level of the perturbation theory, frequently employed to calculate the 
		potential energy curves (PECs)
		of the zero-range model, is based on the diagonalization of the corresponding Hamiltonian in the finite basis 
		set of the unperturbed atomic Rydberg states. This approach is associated with the issue that the eigenenergies 
		do not converge with the increasing number of the Rydberg manifolds included in the basis set \cite{Fey2015}. 
		Another complication inherent to this method is the selection of the kinetic energies at which the scattering 
		lengths are taken. The classical kinetic energy of the Rydberg electron at the position of the perturber, that 
		determines their interaction in the zero-range model, depends on the total energy of the molecular electronic 
		state. This is, however, not known at the stage of the computation when the Hamiltonian matrix is constructed. 
		An alternative approach to the direct diagonalization of the zero-range Hamiltonian in the finite basis set of 
		the atomic Rydberg states is the solution of the integral equation involving corresponding Green's function 
		\cite{Fey2015}.
		
		The first study where the neutral perturber was represented by a finite-range potential, was 
		due to \citet{Khuskivadze}. This interaction was, similarly to the zero-range Fermi's pseudopotential, 
		optimized to correctly reproduce the phase shifts of the electron collisions with the atomic perturber.
		
		The most frequently utilized experimental technique to prepare the LRRMs is their photoassociation in the 
		ultra-cold atomic ensemble via excitation of the atoms into the Rydberg states (see the review by 
		\citet{Shaffer2018} and references therein). The involved 
		atomic states typically possess the principal quantum numbers $n>30$ and corresponding relevant 
		internuclear distances lie above 200 a.u. \cite{Shaffer2018} where the Fermi's zero-range model provides 
		quantitatively satisfactory accuracy of the calculated electronic and vibrational energies.
		
		However, \citet{Bellos2013} carried out an experiment in which different mechanism was utilized to prepare the 
		LRRMs \rbtwo: First, weakly bound \rbtwo 
		molecules in their lowest excited electronic and high vibrational state $\Ket{a^3\Sigma_u^+,\,\nu=35}$ were 
		photo-associated in the magneto-optical trap containing ultra-cold rubidium atoms. The LRRMs with the energies 
		between the states $5s+7p$ and $5s+12p$ (in the asymptotic limit of the 
		separated atoms) were then directly photo-excited. The information about the populations of different 
		electronic and vibrational energy levels of the LRRMs was retrieved using the autoionization spectroscopy of 
		the molecular cations \rbtwoplus. Later, the states slightly red-shifted with respect to the $5s+7p$ energy 
		were studied in more detail by \citet{Carollo2017}.
		
		The outer vibrational turning point of the initial molecular state $\Ket{a^3\Sigma_u^+,\,\nu=35}$ is 
		approximately at the internuclear separation $\approx35$ a.u. and relatively low Rydberg energies are populated 
		by the subsequent photo-excitation. As a result, the LRRMs prepared by \citet{Bellos2013} possess significantly 
		lower 
		internuclear separations than those produced via the direct photo-association of the atomic pair utilized in 
		majority of other experiments.
		
		Although \citet{Bellos2013} successfully associated some of the features observed in their experimental spectra 
		with the structures in the PECs calculated using the Fermi's pseudo-potentials 
		\cite{Greene-prl,Hamilton2002}, 
		the 
		validity of this model becomes questionable at these small internuclear separations and low energies. Since the 
		size of the perturber is not negligible compared to  its distance from the Rydberg atomic core, their mutual 
		interaction may become relevant. Similarly, the size of the perturber is not negligible compared to the de 
		Broglie wave length of the Rydberg electron.
		
		In order to address the validity of the zero-range model of the LRRMs at small internuclear distances, this 
		article is dealing with the PECs of \rbtwo calculated using different models. The calculations are focused on 
		the intermediate range of the internuclear separations below 
		200 a.u., similar to that studied by \citet{Bellos2013}. A more advanced approach, suitable for this range of 
		the 
		nuclear distances, where the valence 
		electron of the alkali-metal atomic perturber is explicitly represented as well as the Rydberg electron, was 
		formulated by \citet{Tarana2016} (hereafter referred to as TC). The construction of this model for \rbtwo is 
		discussed in the present 
		article along with the comparison between the obtained PECs and those published by \citet{Bellos2013} to 
		establish the range of the internuclear distances where the zero-range model provides quantitatively accurate 
		results.
		
		The two-electron \rmat approach does not utilize any external parametrization of the interaction between the 
		Rydberg electron and neutral perturber that depends on the kinetic energy of the electron. Instead, the 
		solutions of the two-electron \schr equation are calculated in the vicinity of the perturber affected by the 
		positive Rydberg core. These are, in terms of the logarithmic derivative, smoothly matched to the wave 
		functions 
		calculated farther from the perturber that satisfy the bound-state asymptotic boundary conditions.
		
		This paper presents the first application of the two-electron \rmat approach \cite{Tarana2016} to 
		other molecular system than H$_2$. Another goal of this work is to present the PECs associated with the 
		perturber in its excited state as these exist below the ionization energy in \rbtwo and can be calculated using 
		the approach developed in TC.
		
		The probability densities of the Rydberg electron are also presented in terms of a one-particle finite-range 
		model of the LRRMs \cite{Khuskivadze} in order to understand their features that are specific to the range of 
		the distances between the nuclei studied in this work.
		
		The spin-orbit couplings and spin-spin couplings are not considered in the calculations presented here. 
		Although their effects can be experimentally recognized in the heavy alkali metals 
		\cite{Deiss2020,Shaffer2018,Eiles2019}, the phenomena investigated in this article are not directly related to 
		the relativistic effects. Adding corresponding degrees of freedom to the two-electron \rmat method formulated 
		in TC would, for the study presented here, yield computationally more demanding calculations without providing 
		equivalent additional insight into the underlying mechanisms.
		
		The rest of this article is organized as follows: The essentials of the two-electron \rmat theory of the LRRMs 
		are reviewed in \secab\ref{sec:theosummary}. Sections~\ref{sec:numasp} and~\ref{sec:onel} are 
		dealing with the parameters of the two-electron and one-electron \rmat calculations, 
		respectively. The results of the calculations are analyzed in Sections \ref{sec:pecs} and \ref{sec:relation}. 
		The conclusions are formulated in \secab\ref{sec:conc}. The model potential of \rbplus and corresponding 
		quantum defects are discussed in Supplemental Material. 
		
		Unless explicitly stated otherwise, the atomic units are used throughout the rest of this article.
	\section{Summary of two-electron \rmat method}
		\label{sec:theosummary}
		Only the essential elements of the two-electron \rmat method are summarized in this section. For its detailed 
		description, an interested reader is referred to the paper TC. Following the notation used in TC, the core of 
		the perturber and the Rydberg core are denoted by \cora and \corb, respectively.
		
		The Rydberg electron is explicitly represented in the model as well as the valence electron of the perturber. 
		The center of the coordinate system is located on the nucleus of the perturber, the Rydberg core is located on 
		the $z$-axis. The coordinates of the electrons are denoted by $\mathbf{r}_1$ and 
		$\mathbf{r}_2$, the position of the positive Rydberg core \corb is denoted by $\mathbf{R}$ ($R$ is the 
		internuclear separation). The radial 
		coordinate of $\mathbf{r}_i$ is $r_i$ and $\Omega_i$ is used for the corresponding angular component.
		
		The Hamiltonian of this system is
		\begin{equation}
			\hat{H}=\sum_{p=1,2}\left[\hat{K}_p+V_A(r_p)+V_B(\left|\mathbf{r}_p-\mathbf{R}\right|)\right]+\frac{1}{r_{12}},
			\label{eq:fullham}
		\end{equation}
		where $\hat{K}_p=-(1/2)\Delta$ is the kinetic-energy operator of the $p$-th electron, $V_A(r)$ and $V_B(r)$ 
		are the potentials representing the positive cores \cora and \corb, 
		respectively. The last term is the repulsion between the Rydberg and valence electron. The interaction of the 
		nuclei $1/R$ is a constant that is added to the final calculated energies and it skipped from all the equations 
		for the brevity of the notation. Corresponding 
		time-independent \schr equation for a fixed internuclear distance $R$ reads
		\begin{equation}
			\hat{H}\Psi(\mathbf{r}_1,\mathbf{r}_2)=E\Psi(\mathbf{r}_1,\mathbf{r}_2),
			\label{eq:scheq}
		\end{equation}
		where $\Psi(\mathbf{r}_1,\mathbf{r}_2)$ is the two-electron bound-state wave function and $E$ denotes the 
		corresponding eigenenergy. Projection of the total angular momentum $M=m_1+m_2$ on the nuclear axis is 
		conserved. The projections of the single-electron angular momenta on the nuclear axis are denoted by $m_1$ and 
		$m_2$.
		
		The valence electron is located exclusively in the vicinity of the perturber core \cora and only 
		the Rydberg electron can appear in the remaining space. A sphere $\mathcal{S}$ centered on the core \cora is 
		introduced with radius $r_0<R$ and large enough to confine all the region where the probability density of the 
		valence electron does not vanish. Then it is sufficient to treat the complicated two-electron \schr equation 
		only in the 
		relatively small inner region and to smoothly match its solutions on the sphere to the single-particle wave 
		functions calculated in the outer region.
		\subsection{Outer region}
		\label{sec:outertheo}
		When the Rydberg electron is located in the outer region, the solution 
		$\left.\Psi(\mathbf{r}_1,\mathbf{r}_2)\right|_{r_2\geq r_0}$ can be expressed using the bound states 
		$\varphi_{im_1}(\mathbf{r}_1)$ of the valence electron weakly affected by 
		the Coulomb tail of $V_B(\left|\mathbf{r}_1-\mathbf{R}\right|)$ defined by the equation
		\begin{subequations}
			\label{eq:perbs}
			\begin{eqnarray}
				\left[\hat{K}_1+V_A(r_1)+V_B(\left|\mathbf{r}_1-\mathbf{R}\right|)\right]&\varphi_{im_1}(\mathbf{r}_1)&\nonumber\\
				=\epsilon_{im_1}&\varphi_{im_1}(\mathbf{r}_1)&,\label{eq:pertbseq}\\
				\left.\varphi_{im_1}(\mathbf{r}_1)\right|_{r_1\geq r_0}\equiv0,\label{eq:pertbsbc}
			\end{eqnarray}
		\end{subequations}
		where $i$ indexes the bound states of the valence electron and $\epsilon_{im_1}$ are the corresponding discrete 
		energies. Using these 
		states, the total two-electron wave function $\left.\Psi(\mathbf{r}_1,\mathbf{r}_2)\right|_{r_2\geq r_0}$ 
		can be expressed as
		\begin{equation}
			\left.\Psi(\mathbf{r}_1,\mathbf{r}_2)\right|_{r_2\geq 
			r_0}=\sum_{im_1}\varphi_{im_1}(\mathbf{r}_1)X_{im_1}(\mathbf{r}_2).
			\label{eq:outerwfex}
		\end{equation}
		Projection of the \schr equation \eqref{eq:scheq} on the states $\varphi_{im_1}(\mathbf{r}_1)$, along with the 
		assumption that the interaction between the Rydberg electron and the perturber can be neglected in the outer 
		region $\left[V_A(r\geq r_0)\equiv0\right]$, yields the following uncoupled set of the equations:
		\begin{equation}
			\left[\hat{K}_2+V_B(\left|\mathbf{r}_2-\mathbf{R}\right|)-(E-\epsilon_{im_1})\right]X_{im_1}(\mathbf{r}_2)=0.
			\label{eq:chaneqs}
		\end{equation}
		Introducing the spherical 
		harmonics $Y_{lm}(\Omega_2)$ on the unit sphere centered on the core \cora, $X_{im_1}(\mathbf{r}_2)$ can be 
		expanded as
		\begin{equation}
			X_{im_1}(\mathbf{r}_2)=\sum_{\substack{l_2\\m_2=M-m_1}}^\infty
			\frac{x_{\overline{j}}(r_2)}{r_2}Y_{l_2m_2}(\Omega_2),
			\label{eq:xpwex}
		\end{equation}
		where the multi-index $\overline{j}=\{i,m_1,l_2\}$.
		The radial wave functions $x_{\overline{j}}(r_2)$ satisfying the asymptotic bound-state boundary conditions 	
		can be obtained using the Green's function $G_{im_1}(\mathbf{r}_2,\mathbf{r}_2')$ for the interaction 
		$V_B(r_2)$ defined by the equation
		\begin{eqnarray}
			\left[\hat{K}_2+V_B(\left|\mathbf{r}_2-\mathbf{R}\right|)-(E-\epsilon_{im_1})\right]
			G_{im_1}(\mathbf{r}_2,\mathbf{r}_2')\nonumber\\
			=-\delta^3(\mathbf{r}_2-\mathbf{r}_2')
			\label{eq:gfdef}
		\end{eqnarray}
		with the following expansion in terms of the spherical harmonics:
		\begin{equation}
			G_{im_1}(\mathbf{r}_2,\mathbf{r}_2')=\sum_{l_2m_2l_2'}\frac{g_{\overline{j}l_2'm_2}(r_2,r_2')}{r_2r_2'}
			Y_{l_2m_2}^*(\Omega_2)Y_{l_2'm_2}(\Omega_2').
			\label{eq:gfpwex}
		\end{equation}
		Subtraction of \eqab\eqref{eq:chaneqs} multiplied by $G_{im_1}(\mathbf{r}_2,\mathbf{r}_2')$ from 
		\eqab\eqref{eq:gfdef} multiplied by $X_{im_1}(\mathbf{r}_2)$, integration throughout the whole outer region in 
		the variable $\mathbf{r}_2$ and evaluation on the sphere $\mathcal{S}$ using the partial-wave 
		expansions \eqref{eq:xpwex} and \eqref{eq:gfpwex} yields the following relation between the 
		values $x_{\overline{j}}(r_0)$ and the radial derivatives of $x_{\overline{j}}'(r_0)$ of the radial wave 
		functions on the sphere:
		\begin{equation}
			\sum_{l_2\in\overline{j}}\left[\Gamma_{\overline{j}l_2}x_{\overline{j}}'(r_0)
			-\Gamma_{\overline{j}l_2}'x_{\overline{j}}(r_0)\right]=0,
			\label{eq:outerbcond}
		\end{equation}
		where $\Gamma_{\overline{j}l_2}=g_{\overline{j}l_2m_2}(r_0,r_0)$, the index $m_2$ was dropped from 
		$\Gamma_{\overline{j}l_2}$ for the brevity of the notation as $m_2=M-m_1$. The radial derivative 
		of the Green's function on the sphere is defined as
		\begin{equation}
			\Gamma_{\overline{j}l_2}'=g_{\overline{j}l_2}'(r_0,r_0)=\lim_{r_2'\to r_0^-}\left.\frac{\partial}{\partial 
			r}\right|_{r_2=r_0}g_{\overline{j}l_2}(r_2,r_2').
		\end{equation}
		Equation \eqref{eq:outerbcond} holds for 
		the wave functions of the Rydberg electron on the sphere that vanish asymptotically for an arbitrary negative 
		value of $E$. The discrete energies of the bound states are determined by the condition that the inner-region 
		wave functions are required to satisfy 
		\eqab\eqref{eq:outerbcond} as well. This selects the energies at which the solutions of the \schr equation 
		\eqref{eq:scheq} are continuous everywhere in the 
		space and satisfy the asymptotic bound-state boundary conditions.

		The Green's function $G_{im_1}(\mathbf{r}_2,\mathbf{r}_2')$ is the Coulomb Green's function \cite{Hostler1963} 
		with 
		the correction for the short-range interaction
		introduced by \citet{Davydkin}, parametrized by the quantum defects. 
		Therefore, $V_B(\left|\mathbf{r}-\mathbf{R}\right|)$ does not explicitly appear in the outer-region 
		calculations.
		
		The treatment of the outer region discussed above is conceptually identical to that used by 
		\citet{Khuskivadze}. Technically, 
		the ground and excited bound states of the valence electron are in this work explicitly considered in the inner 
		region and this additional degree of freedom is taken into account also in the outer-region equations presented 
		above.
		\subsection{Inner region and \rsmat}
		The relation between the values and radial derivatives of the wave functions on the sphere calculated from the 
		inner region, necessary for \eqab\eqref{eq:outerbcond}, is expressed by the \rsmat. Since the eigenstates of 
		the valence electron 
		$\varphi_{im_1}(\mathbf{r}_1)$ vanish on the sphere, the 
		solutions of the \schr equation \eqref{eq:scheq} calculated for the Rydberg electron in the inner region can be 
		on the \rmat sphere expanded as \cite{Aymar1996}
		\begin{equation}
			\left.\Psi_\beta(\mathbf{r}_1,\mathbf{r}_2)\right|_{r_2=r_0}=\sum_{\substack{im_1l_2\\m_2=M-m_1}}
				\varphi_{im_1}(\mathbf{r}_1)\frac{q_{\overline{j}\beta}(r_0)}{r_0}Y_{l_2m_2}(\Omega_2),
			\label{eq:innerwfex}
		\end{equation}
		where $\beta$ indexes the linearly independent solutions in the inner region corresponding to the same energy 
		$E$. Then the \rsmat is defined as \cite{Aymar1996}
		\begin{equation}
			q_{\overline{j}\beta}(r_0)=\sum_{\overline{j}'}R_{\overline{j}\overline{j}'}q_{\overline{j}'\beta}'(r_0).
			\label{eq:rmatdef}
		\end{equation}
		When a linear combination of the general solutions
		\begin{equation}
			x_{\overline{j}}(r_0)=\sum_{\beta}A_\beta q_{\overline{j}\beta}(r_0)
			\label{eq:lincomb}
		\end{equation}
		satisfying also \eqab\eqref{eq:outerbcond} exists, corresponding energy $E$ is the eigenenergy of the bound 
		state.  Substitution of \eqsab\eqref{eq:rmatdef} and 
		\eqref{eq:lincomb} into \eqab\eqref{eq:outerbcond} yields a homogeneous system of linear algebraic equations. 
		The energies at which a non-trivial solution of this linear system exists are identified as the bound-state 
		eigenenergies. Defining the matrix
		\begin{equation}
			\underline{M}=\underline{\Gamma}-\underline{\Gamma}'\underline{R},
			\label{eq:mmatdef}
		\end{equation}
		the condition for the energy of the bound state can be formulated as $\det(\underline{M})=0$.
		
		The \rsmat \eqref{eq:rmatdef} is in this work calculated by a single diagonalization of the modified 
		Hamiltonian \cite{Robicheaux1991}. First, the Bloch operator $\hat{L}_B$ \cite{Tarana2016,Aymar1996}
		is added to the Hamiltonian \eqref{eq:fullham}. 
		The resulting operator $\hat{H}'$ is hermitian inside the \rmat sphere. 
		Its matrix representation $\underline{H}'$ is constructed using the set of two-electron basis 
		functions restricted to the inner region and antisymmetric with respect to the mutual exchange of the 
		electrons. Their angular components are the spherical harmonics coupled to form the 
		eigenstates of the total angular momentum $L$, its projection on the nuclear axis $M$ and total spin $S$ of the 
		spherical two-electron system. The only element in 
		the Hamiltonian \eqref{eq:fullham} that is not spherically symmetric in the selected coordinate system is the 
		potential $V_B$. Inside the \rmat sphere, it is a tail of the off-center Coulomb potential that is expressed as 
		the multipole expansion and it couples the basis functions with different total angular momenta $L$.  
		The radial components of the two-electron basis functions
		are the \bspls used to represent the open and closed functions that are non-zero and vanishing on the sphere, 
		respectively.
		
		The diagonalization of $\underline{H}'$ yields a set of the real eigenvalues $E_k$ and corresponding  
		eigenstates. The projections of the latter on the sphere yields the surface amplitudes $w_{\overline{j}k}$ that 
		can be used to explicitly construct the energy-dependent \rsmat as a pole expansion 
		\cite{Robicheaux1991,tennyson-rev}
		\begin{equation}
			R_{\overline{j}\overline{j}'}(E)=\frac{1}{2}\sum_k\frac{w_{\overline{j}k}w_{\overline{j}'k}}{E_k-E}.
			\label{eq:weexpand}
		\end{equation}
		The benefit of this approach is that the computationally demanding treatment of the two-particle \schr 
		equation in the inner region is not performed for every energy of the interest. Instead, $\underline{H}'$ is 
		diagonalized only once for every internuclear distance, the \rmat poles and amplitudes are obtained and the 
		matrix $\underline{M}$ is numerically constructed on the energy grid using the explicit formula for the \rsmat 
		\eqref{eq:weexpand}.
		
		Note that this computational method does not explicitly involve the classical local kinetic energy of the 
		Rydberg electron at the position of the perturber that is an essential quantity in the zero-range approach. As 
		a result, in the approach discussed above, the internuclear distances at which the classical kinetic energy of 
		the Rydberg electron is negative does not require different treatment from those where it is positive.
	\section{Parameters of two-electron \rmat calculations}
		\label{sec:numasp}
		The representation of the perturber with the core \cora, including the details of the potential $V_A(r)$, as 
		well 
		as the quantum 
		defects of the Rydberg center \corb, are discussed in \secab I of Supplemental Material.
		
		In order to treat the polarization effects between the Rydberg electron and the valence electron of the neutral 
		perturber accurately, the radius of the \rmat sphere centered at the core \cora was set to $r_0=30$ a.u. The 
		formulation of the \rmat method for the LRRMs \cite{Tarana2016} assumes that the 
		polarization potential of the perturber can be neglected outside the \rmat sphere. Therefore, larger $r_0$ 
		allows for larger portion of the polarization potential included in the calculation.
		
		It is well known that the accurate treatment of the electron interactions with the alkali metals at very low 
		energies, that are relevant in this study, requires to propagate the wave function in the polarization 
		potential to very large distances~\cite{Eiles2018}. In order to converge the phase shifts at the low energies, 
		thousands of atomic units are typically necessary \cite{Eiles2018,Tarana2019}. This is due to their large 
		polarizability and the polarization potential affecting the phase of the scattering wave function even at large 
		distances from the target. However, in the LRRMs, except relatively small 
		vicinity of the perturber, it is the Coulomb tail of $V_B(\left|\mathbf{r}-\mathbf{R}\right|)$ 
		that dominates over the polarization potential of the perturber and its effect is treated accurately by the 
		Coulomb Green's function \cite{Hostler1963,Khuskivadze,Tarana2016}.
		Therefore, the calculations of the LRRMs PECs performed with $r_0=30$ a.u. yield very accurate energies 
		although the same radius, while neglecting the polarization potential outside the \rmat sphere, would not 
		provide accurate scattering phase shifts at low energies. 
		
		The angular basis set used  inside the \rmat box consists of the eigenstates $\mathcal{Y}_{l_1l_2}^{(LM)}$ of 
		the 
		total two-electron angular momentum with quantum numbers $L$ and $M$. The one-particle angular momenta of the 
		individual electrons are denoted as $l_1$ and $l_2$. A wider range of the angular momenta was included in the 
		basis set for the calculation performed at smaller internuclear separation $R$ than for those at bigger $R$. 
		The 
		reason is that the Coulomb tail of $V_B$ that breaks the overall spherical symmetry of the two-electron system, 
		varies within the \rmat sphere more rapidly at smaller $R$. As a result, components of the wave function with 
		higher angular momenta are required to converge the calculation inside the \rmat sphere. The basis functions 
		with $l_{1,2}\leq5$ were 
		used for $R\geq70$ a.u. For $70\text{ a.u.}>R\geq55$ a.u., the angular basis set was extended to 
		$l_{1,2}\leq6$, 
		for $55\text{ a.u.}>R\geq42$ a.u. to $l_{1,2}\leq7$ and to $l_{1,2}\leq8$ for even smaller internuclear 
		distances. All the elements with $L$ where $\left|l_1-l_2\right|\leq L\leq l_1+l_2$ were included in the basis 
		set and corresponding blocks coupled by $V_B$ were included in the Hamiltonian matrix $\underline{H}'$.
		
		Every extension of the angular space significantly 
		increases the size of the Hamiltonian matrix $\underline{H}'$. This raises the issues with the computer memory 
		and time necessary for the construction and diagonalization of $\underline{H}'$. In order to keep the 
		calculations computationally 
		tractable, the angular space was extended only at smaller values of $R$ where it is necessary.
		
		The high computational demands of inner-region calculations for small $R$ 
		restrict the research of the PECs to $R\geq37$ a.u. More fundamental lower limit of the internuclear distances 
		at which the \rmat method can be applied, is the radius of the \rmat sphere $r_0$. The presence of the positive 
		core \corb inside the \rmat sphere would require its different representation and consequently reformulation of the inner-region treatment.
		
		As it is discussed in TC, the two-electron configurations in the close-coupling expansion of the wave function 
		inside the \rmat sphere involve the closed one-particle orbitals that vanish on the \rmat sphere -- the lowest 
		bound states of the bare perturber, and the open one-electron orbitals represented by all the 
		\bspls. In this study, the closed orbitals $5s$, $5p$, $4d$, $6s$ and $6p$ were used.
		
		Four lowest eigenstates $\varphi_{im_1}(\mathbf{r}_1)$ of the perturber in the Coulomb tail of the potential 
		$V_B(\left|\mathbf{r}-\mathbf{R}\right|)$ were included in the calculations as the scattering channels [see 
		\eqsab\eqref{eq:perbs}, \eqref{eq:outerwfex} and \eqref{eq:innerwfex}]. They correspond to the ground state 
		$5s$ of Rb and the excited state $5p$ split by the non-spherical off-center Coulomb potential $V_B$. They 
		are essential for the calculation of the PECs involving the excited states of the perturber.
	\section{One-electron \rmat calculations}
		\label{sec:onel}
		The two-particle character of the electronic wave function inside the \rmat sphere taken into account in 
		the two-electron \rmat method \cite{Tarana2016} limits its possibilities to visualize this wave function. 
		Although its single-particle component in the outer region, that is most important for this study, can be 
		plotted easily, that information is not sufficient at the energies and internuclear distances 
		discussed in this paper as the size of the \rmat sphere is comparable with the size of the overall region where 
		the Rydberg electron can be located.
		
		Another motivation for these finite-range single-particle calculations is the fact that the energies obtained 
		within the zero-range model can be sensitive to its particular computational 
		implementation (direct diagonalization in the finite basis set or calculations directly involving the Green's 
		functions) and to the parameters of the particular calculation, i.e. the number of the basis functions involved 
		in the diagonalization \cite{Fey2015}. However, the references \cite{Bellos2013,Carollo2017} do not include a 
		detailed 
		discussion of the numerical parameters used to obtain the results. Therefore, their correspondence with the 
		finite-range single particle PECs presented in this study, in addition to the agreement with the experimental 
		spectra, further supports their accuracy. Both approaches represent different technical realizations of similar 
		physical approximations -- model-potential \electron-Rb interaction that yields the correct scattering phase 
		shifts.
		
		In order to investigate the features of the Rydberg wave functions, another single-particle model of the LRRMs 
		was utilized where the perturber was represented by a finite-range potential. Although this model is more 
		approximate than the two-electron treatment, the PECs obtained from both approaches (discussed in 
		\secab\ref{sec:pecs}) are, except several specific regions, in good qualitative and quantitative agreement.
		
		The finite-range potential representing the neutral rubidium perturber in the single-particle part of this 
		study was constructed by 
		\citet{Khuskivadze} and successfully utilized in the context of the LRRMs \cite{Khuskivadze} as well as the 
		near-threshold photodetachment of the alkali metals anions \cite{Bahrim2002}. It was optimized to accurately 
		reproduce the low-energy phase shifts of the \electron-Rb collisions (the same condition as in the case of the 
		zero-range potentials -- see \cite{Eiles2019} and the references therein). Note, however, that the potential 
		itself is energy-independent and it is not a function of the kinetic energy of the electron. The \electron-Rb 
		interaction is considered only in the partial waves $s$ and $p$.
		
		Although \citet{Khuskivadze} also included the spin-orbit term in their potential, it was disregarded in 
		this work in order to construct the model that is consistent with the two-electron approach and only the 
		triplet component was used.
		
		The computational method used to obtain the PECs of the LRRMs is also 
		similar to that of \citet{Khuskivadze}. The configuration space of the Rydberg electron is separated by a 
		sphere centered on the perturber and sufficiently large so that the interaction of the Rydberg electron with 
		the perturber can be neglected outside. The treatment of the outer region utilized in the single-electron 
		calculations presented here is the same as in the reference \cite{Khuskivadze}. It is a special case of the 
		approach discussed in \secab\ref{sec:outertheo}. The single electron finite-range calculations were performed 
		with the quantum defects of \rbplus published by \citet{Lorenzen1983}, while the two-electron 
		\rmat results are obtained using the quantum defects calculated from the model potential $V_A$ (see \secab I of
		Supplemental Material).
		
		In the calculations presented here, the Schr\"odinger equation inside the sphere was not directly 
		numerically integrated for every energy of the interest as in the reference \cite{Khuskivadze}. Instead, the 
		inner-region problem was formulated in terms of the \rsmat and treated along the same lines as in 
		\secab\ref{sec:theosummary}.
		
		The spherical harmonics $Y_{lm}$ with $l=0\dots9$ and $m=0$ were used as an angular basis in 
		the inner region where the wave function was expanded with respect to the center of the sphere. Although the 
		model potential \cite{Khuskivadze} is non-zero only for $l=0,1$, the off-center Coulomb potential due to the 
		Rydberg core requires the higher angular momenta, particularly at smaller internuclear separations. The radius 
		of the \rmat sphere was in the single-electron calculations set to 30 a.u. The radial part of the inner-region 
		wave function was expressed as a linear combination of 200 \bspls of the 6th order.

	\section{Computed potential energy curves}
		\label{sec:pecs}
		The \tsigma PECs of the long-range Rydberg states in \rbtwo 
		were calculated using the two-electron \rmat method for the internuclear distances $R$ between 37 a.u. and 200 
		a.u. The energies of our interest 
		span the interval between the dissociation thresholds corresponding to the states of the non-interacting atoms 
		$5s+7p$ and $5s+13p$.
		
		The single-electron \rmat calculations based on the finite-range potentials by \citet{Khuskivadze} 
		discussed in \secab\ref{sec:onel} were performed for the same interval of the energies and internuclear 
		distances. Since these potentials were constructed to accurately fit 
		the same low-energy \electron-Rb phase shifts as the contact potentials used by \citet{Bellos2013}, it is not 
		surprising that these two approaches yield quantitatively very similar results. Therefore, the comparison of 
		these single-electron approaches throughout the whole studied range of the energies and $R$ is not presented 
		here. Instead, their similarity is illustrated for the selected states below in this section and the regions 
		where they yield qualitatively different PECs are discussed.
	
		Small segments of several PECs presented below are missing near the energies 
		corresponding to the infinite separation of the nuclei or small discontinuities appear there. It is due to the 
		finite energy grid used in the   numerical matching of the wave functions on the \rmat sphere. Near the 
		asymptotes, the energy grid is not sufficiently fine to separate the molecular states where 
		$\det(\underline{M})=0$ [see \eqab\eqref{eq:mmatdef}] from the energies of unperturbed atomic Rydberg states 
		where the Green's function possesses poles.

		Among the other curves discussed below, the figures with the calculated PECs also show four very 
		steeply rising PECs that do not appear in the results of the zero-range model. They are artifacts of the 
		utilized computational method that are further explained in Appendix.
		
		Note that the two-electron and one-electron finite-range \rmat results presented below take into account the 
		polarization of the neutral perturber by the positive Rydberg core. The inner-region wave function is in the 
		two-electron method expanded in terms of the eigenstates $\varphi_{im_1}$ of the perturber affected by the 
		positive Rydberg center [see \eqsab\eqref{eq:perbs}]. The term
		\begin{equation}
			V_{\mathrm{pol}}=-\frac{\alpha_d}{2R^4},
			\label{eq:polpoten}
		\end{equation}
		where $\alpha_d$ is the static dipole polarizability of rubidium, is included in the finite-range model 
		potential \cite{Khuskivadze} used in the one-electron finite-range model. In order to achieve a like-to-like 
		comparison of these methods with the zero-range model, the term \eqref{eq:polpoten} was also added to all the 
		zero-range energies taken from the reference \cite{Bellos2013} that are presented below. It has been 
		utilized in previously published studies based on the zero-range model to partially represent the 
		interaction between the neutral atom and positive ion 
		\cite{Schlagmuller2016,Schlagmuller2016-prx,Liebisch2016,Whalen2017}.
		\subsection{PECs associated with \rbminus resonance}
\begin{figure}
				\includegraphics{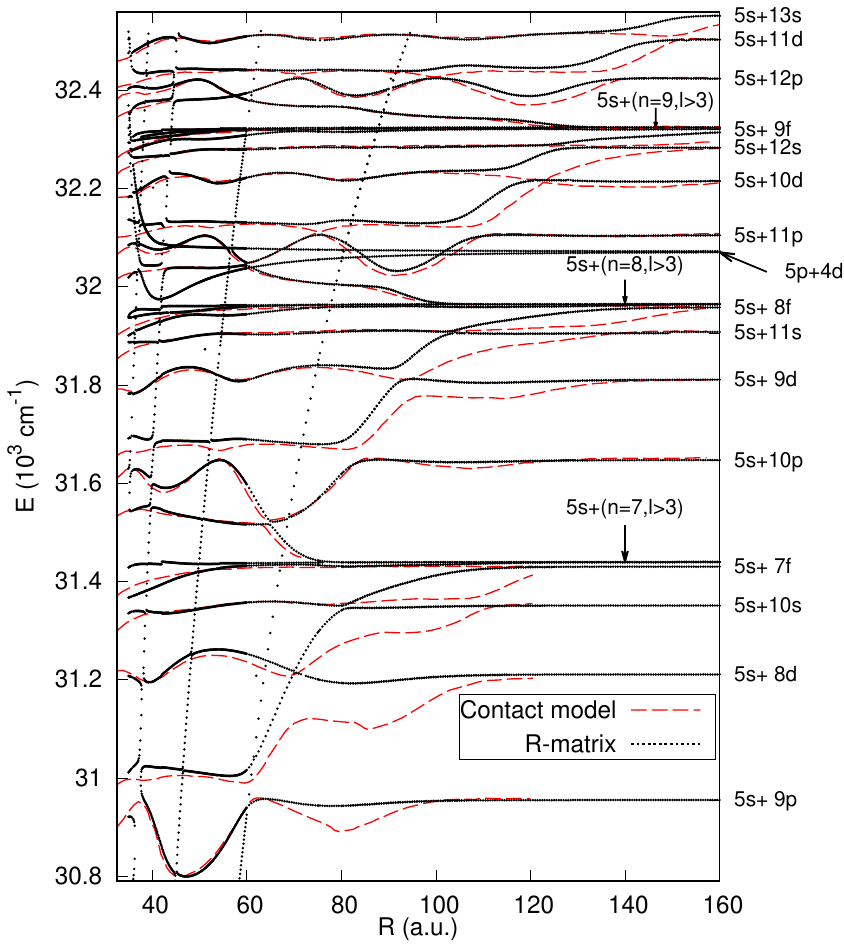}
				\caption{PECs \tsigma calculated using the two-electron \rmat technique (black dotted
				lines) compared with the curves obtained from the zero-range model by \citet{Bellos2013}
				(red dashed lines). Zero energy corresponds to two non-interacting Rb($5s$) atoms. The marks on the 
				right side denote the non-degenerate electronic states of the non-interacting 
				atoms. The vertical arrows denote the degenerate hydrogen-like manifolds of the Rydberg electron $l>3$.}
				\label{fig:lowpecs}
			\end{figure}

\begin{figure}
				\includegraphics{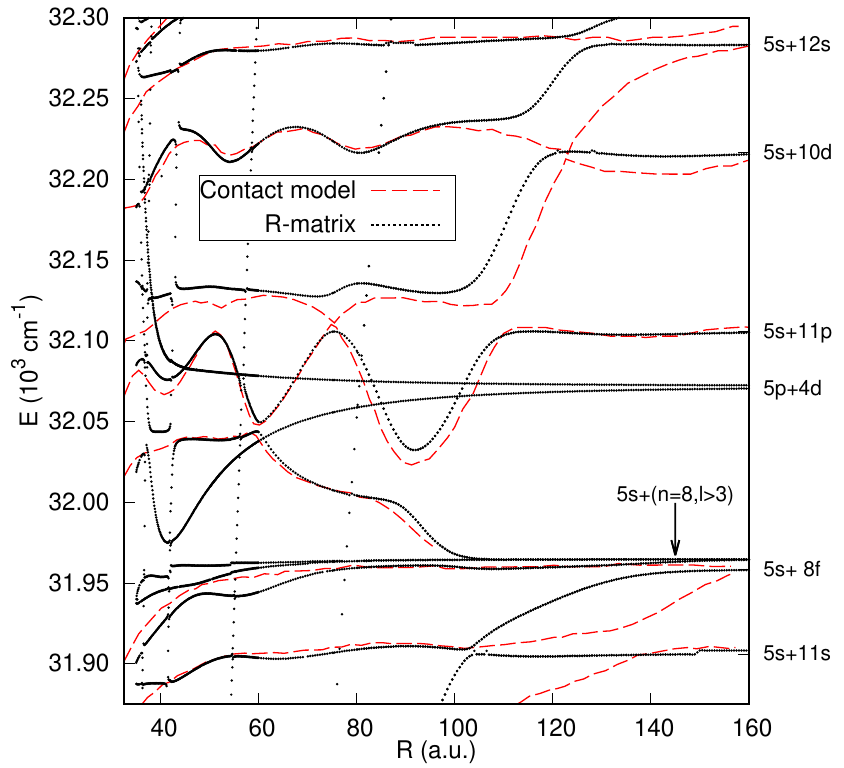}
				\caption{Detail of the \tsigma PECs between the asymptotes $5s+9p$ and $5s+13s$ also shown in 
				\figab\ref{fig:lowpecs}. The notation is identical with \figab\ref{fig:lowpecs}.}
				\label{fig:detail11s12s}
			\end{figure}

			As can be seen in \figab\ref{fig:lowpecs}, the two-electron \rmat method yields qualitatively similar 
			results to those obtained from the zero-range potentials used by \citet{Bellos2013}.
			The quantitative differences are visible for the PECs that detach from the degenerate hydrogenic manifolds 
			$5s+n(l>3)$ ($n=7\dots10$, denoted by the vertical arrows in \figsab\ref{fig:lowpecs} and 
			\ref{fig:detail11s12s}) and descend as $R$ decreases. These curves are a consequence of the low-energy 
			$p$-wave resonance in Rb$^-$ \cite{Bahrim2000,Bahrim2001,Hamilton2002}. The probability density of the 
			Rydberg electron is in the corresponding electronic states localized around the perturber more than in the 
			general molecular Rydberg states.
			
			The zero-range potential is constructed to yield the same phase shifts as the true 
			multi-electron interaction. Its application assumes that the interaction region 
			spanned by the perturber is infinitesimally small, as the parametrization by the phase shifts is used 
			everywhere in the space. On the other hand, the two-electron \rsmat takes into account the full repulsion 
			between 
			the Rydberg electron and the valence electron while their relative distance is small and treats the 
			two-electron wave function in the small interaction region of the perturber where its parametrization by 
			the phase shifts is not accurate. Simultaneously, the influence of the positive core $B$ on both electrons 
			is included as well. These subtle two-electron effects along with the finite size of the 
			perturber can play an important role in the states where the Rydberg electron is predominately localized in 
			the vicinity of the perturber.
			
\begin{figure}
				\includegraphics{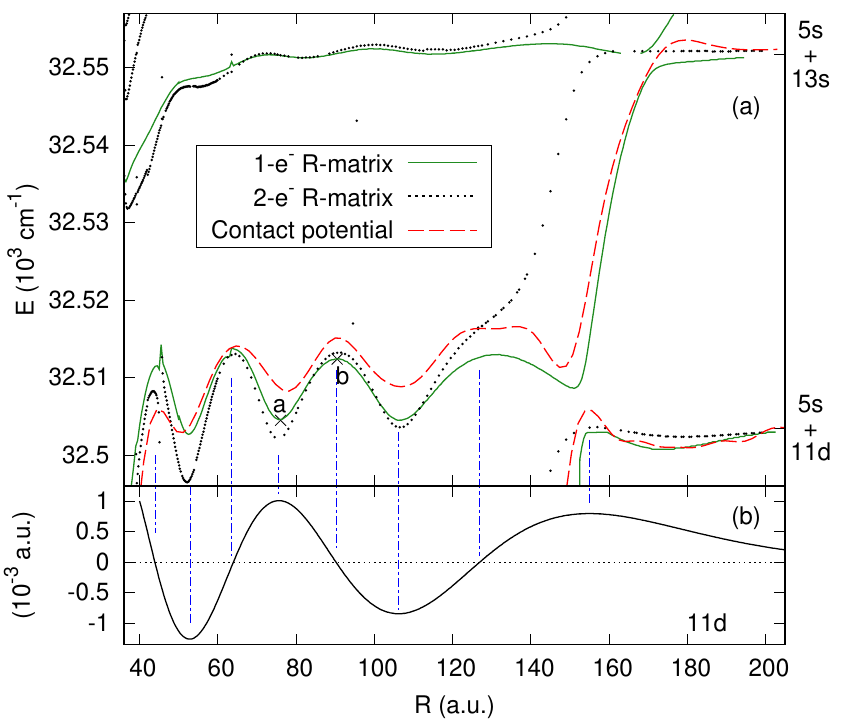}
				\caption{Detail of the PECs with the asymptotes $5s+11d$ and $5s+13s$ calculated using the 
				single-electron and two-electron models (a). The labeled points are those at which the Rydberg electron 
				densities are plotted in \figab 2 of Supplemental Material. Radial values of the atomic Rydberg 
				function $11d$ at the perturber's center (b). The vertical guiding lines connect the structures 
				in the PECs with corresponding extremes and nodes of the atomic Rydberg wave functions.}
				\label{fig:sddetail}
			\end{figure}

			The importance of the interplay between the electron-electron repulsion and effect of the positive Rydberg 
			core 
			treated in the finite volume is also supported by the one-electron \rmat calculations. 
			Figure \ref{fig:sddetail} shows that even the finite-range one-electron approximation of 
			the neutral perturber yields the PECs involving the atomic resonance (the segment steeply descending with 
			decreasing $R$ from the asymptote $5s+13s$) very similar to those obtained from the contact model. This 
			similarity is not affected by the term $\sim-\mathbf{r}.\mathbf{R}/(Rr)^3$ that in the finite-range 
			single-electron model \cite{Khuskivadze} represents the interaction between the Rydberg electron and 
			positive Rydberg core via the dipole moments induced on the perturber. The two-electron wave function and 
			the true Coulomb repulsion of the electrons inside the sufficiently large \rmat sphere are the factors 
			responsible for the differences between the two-electron PECs and those obtained from the single-particle 
			potential models.
			
\begin{figure}
				\includegraphics{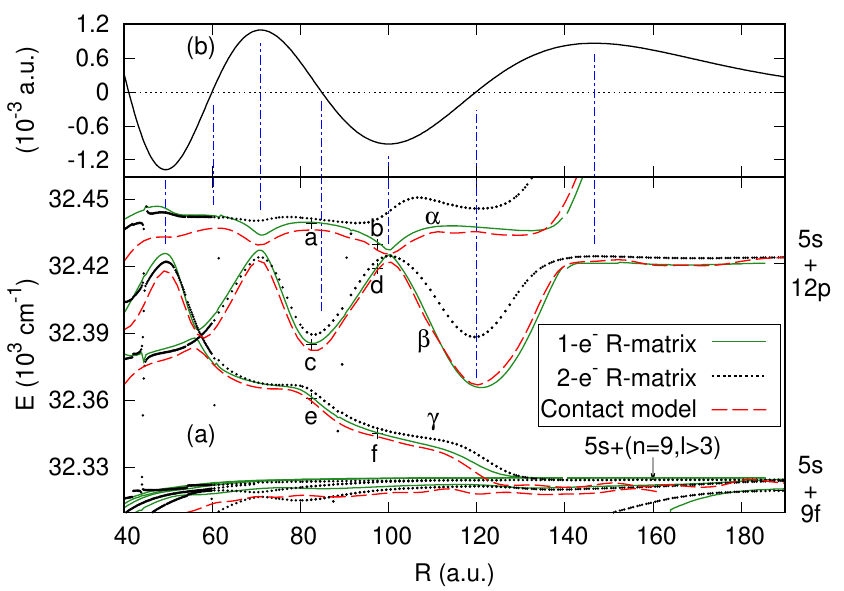}
				\caption{Detail of the PECs with the asymptote $5s+12p$ calculated using the finite-range 
				one-electron representation of the perturber \cite{Khuskivadze} (a). The labeled points are those at 
				which the Rydberg electron densities are plotted in \figab\ref{fig:wfstrongpert}. The radial values of 
				the atomic Rydberg wave function $12p$ plotted at the perturber's center (b). The vertical guiding 
				lines connect the structures in the PECs with corresponding extremes and nodes of the atomic Rydberg 
				wave function.}
				\label{fig:pdetail}	
			\end{figure}

			These factors are also responsible for the 
			disparities between the two-electron and one-electron avoided crossings involving the PECs associated with 
			the \rbminus resonance. Simultaneously, both single-particle approaches yield these avoided 
			crossings very similar to each other for all the states studied here. This is for a subset of the PECs 
			illustrated in \figsab\ref{fig:sddetail} and \ref{fig:pdetail}.
			
		\subsection{Oscillating PECs near asymptotes 5\textit{s}+\textit{np}}
\begin{figure}
				\includegraphics{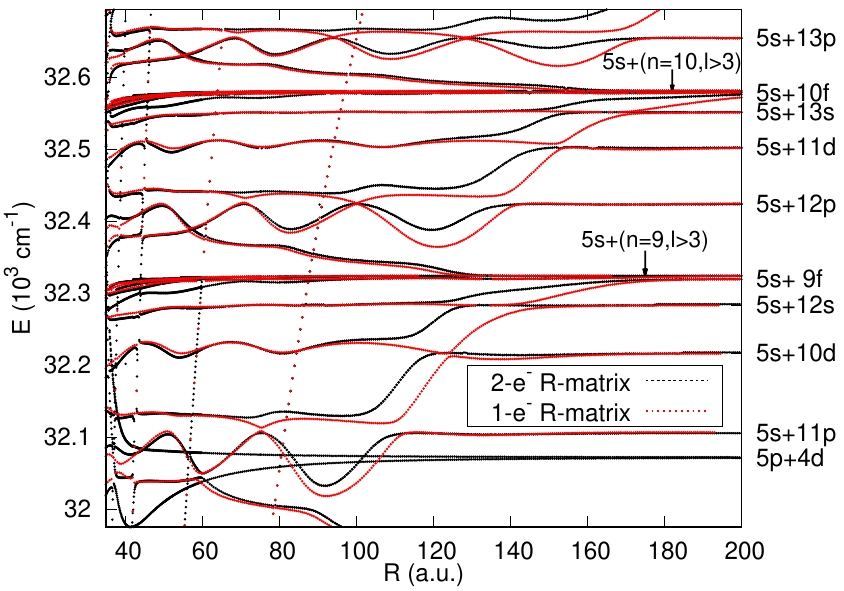}
				\caption{PECs \tsigma calculated using the two-electron \rmat technique (black dotted lines) 
				compared with the curves obtained from the one-electron finite-range calculations \cite{Khuskivadze} 
				(red dotted lines). Zero energy and the meaning of the labels are identical to \figab\ref{fig:lowpecs}.}
				\label{fig:hipecs}
			\end{figure}

			Avoided crossings between the oscillating PECs located at energies closest to the asymptotes $5s+np$ 
			(curves $\alpha$ and $\beta$ in \figab\ref{fig:pdetail} for the asymptote $5s+12p$) also show differences 
			between the single-electron and two-electron approaches. As can be seen in \figab\ref{fig:lowpecs}, 
			\ref{fig:pdetail} and \ref{fig:hipecs} for the asymptotes $5s+np$ where $n=10\dots13$, the 
			avoided crossing left of the outermost deep potential well (near the points b and d in 
			\figab\ref{fig:pdetail} for the asymptote $5s+12p$) obtained from the two-electron \rmat approach is less 
			narrow than the same structure calculated using the finite-range or zero-range single-electron model. The 
			avoided crossings of the same curves at smaller internuclear distances show improving agreement between the 
			single-electron and two-electron calculations with decreasing $R$.
			
			It is not straightforward to attribute these differences to a single effect. Although the 
			probability densities of the Rydberg electron in the states near these avoided 
			crossings are not localized around the perturber as much as in the states involving the \electron--Rb 
			resonance, they are not negligible in this region. Therefore, the two-electron effects can play a role here 
			in a similar way as it is discussed above. On the other hand, the classical 
			kinetic energy of the Rydberg electron at the position of the perturber is in the case of these avoided 
			crossings higher than 100 meV. As a result, the \electron--Rb interaction in the $d$-wave can also 
			partially contribute to the different distances between the avoiding curves. This $d$-wave interaction is 
			included in the two-electron \rmat approach and disregarded in the one-particle models. Its effect was 
			supported by the test two-electron calculation where the interaction of the Rydberg electron with the 
			perturber was artificially restricted only to the partial waves $s$ and $p$. Obtained avoided crossings 
			between the oscillating PECs 
			near the $5s+np$ asymptotes were narrower than those calculated using the full interaction, although not as 
			narrow as the single-particle crossings. These arguments, 
			however, do not explain the improving correspondence among different models for the avoided crossings at 
			smaller $R$.
		\subsection{Classical turning points}
			Both single-electron approaches considered in this work yield, in the region of the energies and 
			internuclear distances discussed above, PECs that are qualitatively and quantitatively similar to each 
			other. This is because both \electron--Rb potentials model the same physical properties of this system -- 
			the $s$-wave and $p$-wave scattering phase shifts.
			
\begin{figure}
				\includegraphics{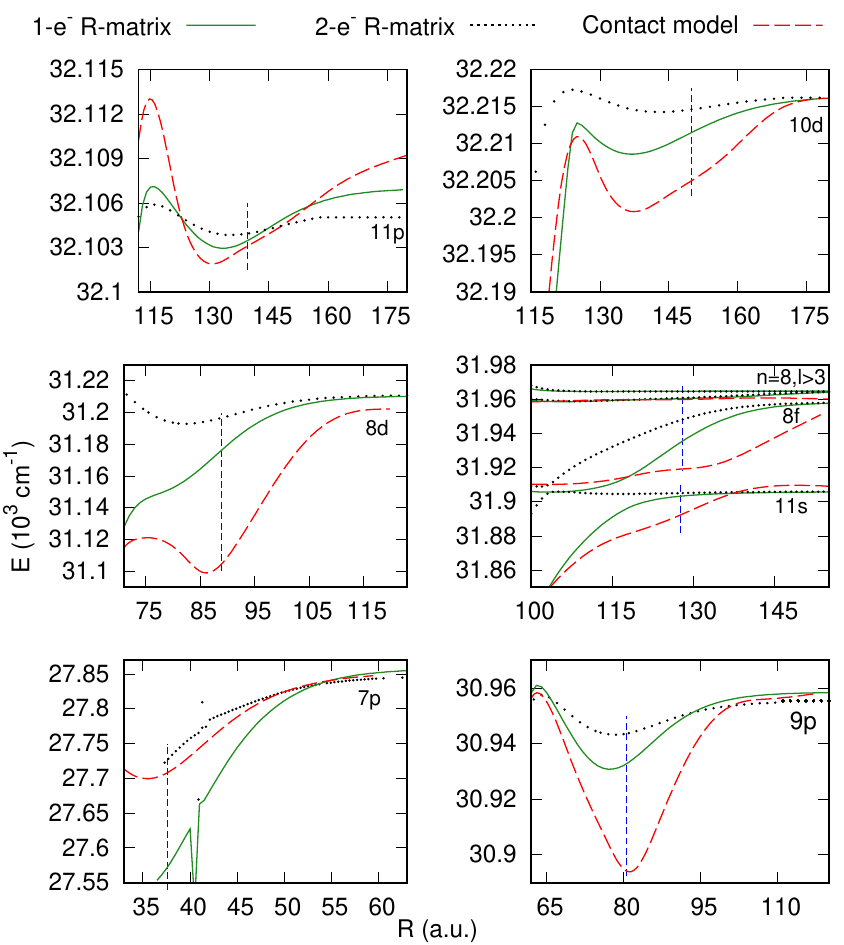}
				\caption{Segments of the PECs calculated using the one-electron and two-electron models where the 
				Rydberg electron is classically forbidden. The vertical blue dashed lines denote the positions of the 
				classical turning points $R_{CT}=-1/\varepsilon_{nl}$ where $\varepsilon_{nl}$ is the energy of the 
				unperturbed atomic Rydberg state $nl$. The variations in the asymptotic energies among compared models 
				are due to different quantum defects used in different models.}
				\label{fig:turnpoints}
			\end{figure}

			This general agreement breaks when the perturber is 
			located in the classically forbidden region of	the Rydberg electron. Application of the zero-range 
			model there requires its extension. To the best of the author's knowledge, the most frequent 
			generalizations utilized in the previously published works are either based on setting the 
			local kinetic energy of the Rydberg electron to zero everywhere in the classically forbidden region 
			\cite{Eiles2019,Anderson} or on the smooth extension of the \electron--Rb phase shifts to the negative 
			energies \cite{Stanojevic2020}. According to \citet{Shaffer2018}, these unphysical approximations  do 
			not have a considerable impact on the nuclear dynamics of the LRRMs discussed there and in the 
			references therein. Application of the zero-range interaction in this region can cause cusps in the 
			calculated PECs at the classical turning points of the Rydberg electron 
			\cite{Eiles2019,Shaffer2018,Anderson}. Although the description of the zero-range potential in the 
			classically forbidden region is not included in the reference \cite{Bellos2013}, these cusps can be clearly 
			recognized for the low-lying PECs with the asymptotes $5s+9p$ and $5s+8d$ in \figsab\ref{fig:lowpecs} and 
			\ref{fig:turnpoints}.
			Their amplitudes decrease with increasing energy and for the higher PECs the cusps cannot be clearly 
			distinguished from the shallow potential wells formed near the classical turning points (see 
			\figab\ref{fig:turnpoints}).
			
			Unlike the contact interaction, the finite-range potential \cite{Khuskivadze} does not directly depend on 
			the local kinetic energy of the Rydberg electron. Consequently, the classical turning points do not appear 
			in the calculation. Technically, the general solutions of the Schr\"odinger equation 
			with this potential can be obtained for arbitrary energies. The bound states are identified with those wave 
			functions at particular 
			negative energies that satisfy the corresponding boundary conditions. As a result, the obtained PECs are 
			smooth near the classical turning points of the Rydberg electron. However, this does not imply that 
			they are more accurate in the classically forbidden regions than the PECs calculated using the zero-range 
			model. Although the local classical kinetic energy of 
			the Rydberg electron does not explicitly appear in the calculations, the model potential was optimized by 
			fitting the \electron--Rb phase shifts only 
			for the positive kinetic energies. Utilization of the same potential in the classically forbidden 
			regions is another arbitrary, although smooth, extrapolation of the model interaction towards the negative 
			local classical kinetic energies of the Rydberg electron.
			
			As can be seen in \figab\ref{fig:turnpoints}, the mutual deviation of the energies obtained from both 
			approaches becomes significant even in the classically allowed region on the length scale similar to the 
			size of the neutral atom. It is a consequence of the finite size of the perturber considered in the model 
			interaction by \citet{Khuskivadze}, 
			
			The two-electron \rmat method is, in the classically forbidden region, free of all the issues associated 
			with 
			the single-electron approaches mentioned above as the \electron--Rb interaction is not constructed by a 
			parametrization of any quantity that depends on the kinetic energy of the impinging electron. Instead, the 
			two-electron wave function 
			inside the \rmat sphere for the Hamiltonian \eqref{eq:fullham} is obtained at an arbitrary energy without 
			imposing any outer-boundary conditions. At the energies of the bound states, it is possible to perform its 
			smooth matching with the asymptotically vanishing outer-region wave function in terms of the \rsmat. This 
			object is not associated with any particular outer-region boundary conditions \cite{Aymar1996} or types of  
			the wave functions in the outer region that depend on the electron kinetic energy. As a result, whole the 
			problem can be solved in terms of the total energy $E$. As can be seen in \secab\ref{sec:theosummary}, it 
			is not necessary to express the kinetic energy of the Rydberg electron anywhere in the calculation. The 
			details of the two-electron PECs in the classically forbidden region are plotted in 
			\figab\ref{fig:turnpoints}.
		\subsection{Short internuclear distances}
			Another region where the PECs obtained from the models discussed here deviate from each other is the range 
			of the small internuclear distances among those studied in this work. An illustration 
			of their discrepancies at small $R$ can be seen in the top PEC visualized in \figab\ref{fig:pdetail} where 
			each single-particle model yields different energies at $R\lesssim60$ a.u. The differences between the 
			one-electron models are due to the fact that at these small distances of the nuclei, the 
			variation of $V_B(\left|\mathbf{r}-\mathbf{R}\right|)$ is not negligible on the scale comparable to the 
			size of the perturber. Therefore, the local kinetic energy of the Rydberg electron is not constant 
			throughout the region occupied by the neutral atom. This effect is not treated by the zero-range model 
			\cite{Bellos2013} and it is taken into account by the finite-range potential \cite{Khuskivadze}.
			
			The differences between the one-particle models and the two-electron \rmat calculations are caused by the 
			polarization of the valence electron of the perturber by both, the Rydberg electron as well as the positive 
			Rydberg core, accurately treated by the two-electron \rmat method [see \eqsab\eqref{eq:perbs}]. The 
			deviations of the 
			PECs can be seen in 
			\figab\ref{fig:lowpecs} for the states with the asymptotes $5s+7f$ and $5s+10p$ at $R\lesssim50$ a.u. and 
			$R\lesssim40$ a.u., respectively. Figure \ref{fig:detail11s12s} provides further 
			illustrations for the state with the asymptote $5s+11p$. The polarization term \eqref{eq:polpoten} in the 
			one-electron approaches improves the overall agreement of the corresponding PECs with the two-particle 
			model at small $R$. It suggests that this simple term reasonably treats the polarization of the perturber 
			by the positive Rydberg core.
			
			Among the other PECs at small internuclear distances mentioned above, the curve with the asymptote $5s+7p$ 
			plotted in \figab\ref{fig:turnpoints} is worth mentioning here since the corresponding classical turning 
			point of the Rydberg electron is located at the small internuclear distance $ R=37.6$ a.u. Beyond this 
			value, the differences among the PECs obtained from the models discussed here are caused by two effects: 
			the polarization of the perturber by the positive Rydberg core as well as by the different treatment of the 
			interaction between the perturber and the Rydberg electron in its classically forbidden region. In the 
			higher PECs mentioned above, each of these effects is dominant in different segment of the curve.
			
			The states near the asymptote $5s+7p$ were also subjects of the experimental study by \citet{Carollo2017}. 
			The energies of the vibrational states supported by this PEC were measured and their good agreement with 
			those calculated from the zero-range PEC was reported.
		\subsection{PECs associated with excited perturber}
			The two-electron \rmat method allows for calculation of the PECs associated with the excited states of the perturber as its valence electron is directly represented. In the 
			limit of infinitely separated nuclei, three dissociation thresholds involving the perturber in its excited 
			state lie below the lowest ionization threshold: $5p+5p$, $5p+4d$ and $5p+6s$. The lowest ($5p+5p$) and 
			highest ($5p+6s$) among them lie below and above the energy interval studied in this article, respectively. 
			The PECs with the asymptotic energy corresponding to 
			the separated atomic states $5p$ and $4d$ are displayed in \figsab\ref{fig:lowpecs} and 
			\ref{fig:detail11s12s}.
			
			Since the excitation energy of the 
			perturber represents here a large portion of the total energy, the Rydberg electron is 
			confined deeper in the potential $V_B$ than in the molecular states in the studied energy range where 
			the perturber is in its ground state. As a result, the potential barrier formed by the Coulomb tail of 
			$V_B$ and polarization potential of the perturber in its excited state is too high for the Rydberg electron 
			to classically reach the  perturber for every internuclear separation beyond $\approx43$ a.u. Therefore, 
			all the segments of the  PECs located at $R\gtrsim43$ a.u. can be very accurately approximated as
			$E(R\gtrsim43\ \text{a.u.})=\epsilon_{2m_1}(R)+\varepsilon_{42}$
			where $\varepsilon_{42}$ is the energy of the atomic Rydberg state $4d$ and 
			$\epsilon_{2m_1}(R)$ are the energies of the perturber in the excited state $5p$ polarized by the Coulomb 
			tail of $V_B$ [see \eqsab\eqref{eq:perbs}]. The triply degenerate state $5p$ of the perturber is split by 
			$V_B$    into the lower state with $m_1=0$ and the higher doubly degenerate state where $m_1=\pm1$. The 
			angular 
			momenta of the Rydberg electron are then coupled to form $\Sigma$-states of the two-electron system.
		
			Validity of this approximation breaks at $R\lesssim43$ a.u. where the lower PEC shows clear minimum and the 
			higher doubly degenerate anti-bonding curve shows even more steep anti-bonding character. Although the 
			lower curve supports the bound vibrational states, their experimental realization or observation might be  
			generally difficult due to low lifetimes of the involved low excited atomic states \cite{Safronova2011}.
		
	\section{Relation between Rydberg wave functions and structures in PECs}
	\label{sec:relation}
	The structures in the calculated PECs are related to the character of corresponding electronic wave functions. 
	Several categories of the LRRM states, that have previously been well studied at larger nuclear distances, show 
	different character in the range of smaller $R$ studied in this work.

	Since the two-electron wave functions obtained from the \rmat calculations are complicated 
	to analyze, especially inside the \rmat sphere, the structure of the Rydberg electron wave 
	functions was studied by means of the single-particle finite-range model.
	
	As can be seen in \figab\ref{fig:hipecs}, the one-electron 
	finite-range model yields PECs qualitatively very similar to those 
	obtained from the two-electron approach.	Therefore, it is reasonable to anticipate that 
	the structures found in the corresponding wave functions of the Rydberg electron are also qualitatively correct and 
	allow for the characterization of the two-electron PECs.
	
	The overall qualitative agreement among the PECs calculated using all three models discussed in this article 
	implies that, except small $R\lesssim40$ a.u., the features discussed in the rest of this section are not 
	consequences of the finite size of the perturber. In fact, as it is discussed below, they can be, at least 
	qualitatively, explained in terms of the zero-range model. They are specific for the range of the energies and 
	internuclear separations discussed in this work.

	The PECs with the asymptotes $5s+ns$, $5s+nd$ and $5s+nf$ plotted in \figsab\ref{fig:lowpecs}, 
	\ref{fig:detail11s12s} and \ref{fig:hipecs} possess oscillatory structures. As it is illustrated in 
	\figab\ref{fig:sddetail} for the state with the asymptote $5s+11d$, the positions of the local maxima and minima in 
	these PECs very accurately agree with the locations of the radial nodes and local extremes in corresponding 
	unperturbed atomic Rydberg wave functions, respectively. This is consistent with the perturbative treatment of the 
	zero-range model \cite{Greene-prl} where the negative energy shift is proportional to the probability density of 
	the unperturbed Rydberg electron at the position of the perturber. Corresponding illustrative two-dimensional maps 
	of 
	the probability density of the Rydberg electron are presented in \secab III of Supplemental Material. For the 
	nuclear 
	geometries where the PECs attain their local minima and maxima, the wave functions of the Rydberg electron are in 
	the vicinity of the neutral perturber dominated by the $s$-wave and $p$-wave (with respect to the perturber), 
	respectively.

	\subsection{PECs with asymptotes 5\textit{s}+\textit{np} and 5\textit{s}+\textit{n}(\textit{l$>$}3)}
		The electronic wave functions of the states with the asymptotes $5s+np$ have very different character 
		from those discussed above. As can be seen in \figsab\ref{fig:lowpecs}, 
		\ref{fig:detail11s12s}, \ref{fig:pdetail} and \ref{fig:hipecs}, three curves appear at the energies close to 
		every asymptote $5s+np$: 
		Two of them (denoted as $\alpha$ and $\beta$ in \figab\ref{fig:pdetail}) cross each other and show clear 
		oscillatory behavior. The third curve  (denoted as $\gamma$ in \figab\ref{fig:pdetail}) is 
		monotonically decreasing 
		towards the nearest lower asymptotic degenerate hydrogen-like threshold $5s+(n-3)(l>3)$. Although the 
		discussion 
		in the rest of this subsection is focused on the curves near the threshold $5s+12p$ magnified in 
		\figab\ref{fig:pdetail}, it is valid for all the Rydberg states $np$ considered in this article.
		
		At the internuclear separations near the dissociation limit, the wave function (not visualized here) retains 
		the overall $p$-symmetry (with respect to the Rydberg core) with the perturbation localized in the vicinity of 
		the perturber.
		\begin{figure}
			\includegraphics[width=3.41in]{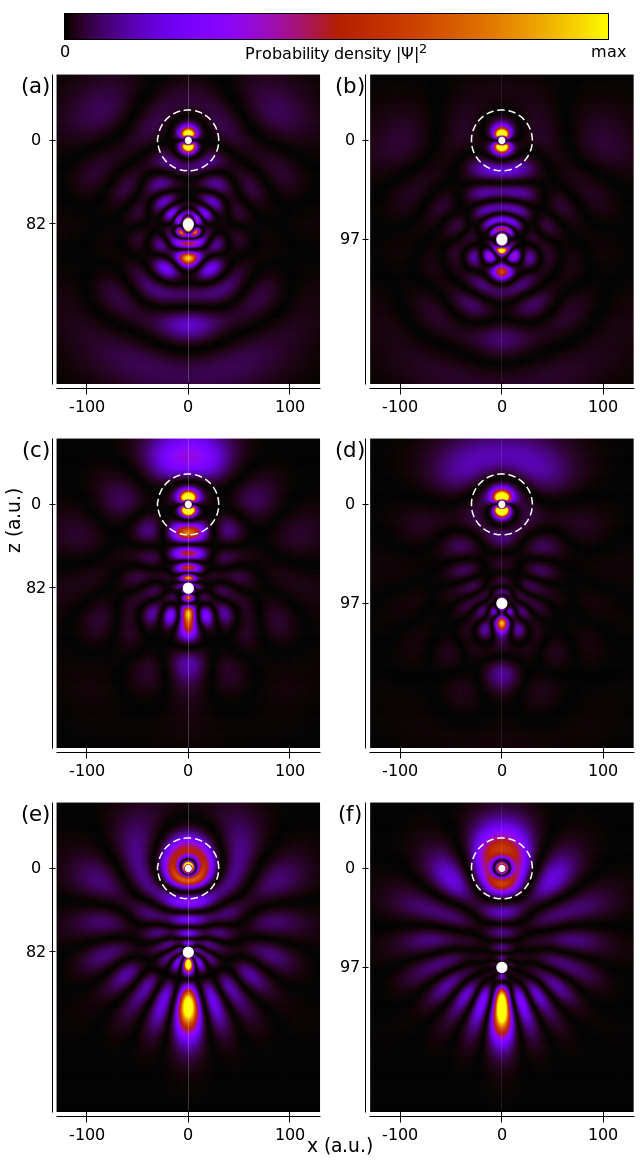}
			\caption{Two-dimensional maps of the Rydberg electron probability densities in the molecular states where 
				the atomic states $12p$ and $10(l>3)$ are perturbed. The perturber is located in the center of the 
				coordinate system and the white circle denotes the \rmat sphere. The internuclear 
				distances are marked on the vertical axes and corresponding points on the PECs are labeled in 
				\figab\ref{fig:pdetail}. The small vicinities of the atomic centers are discussed in Section III of 
				Supplemental Material.}
			\label{fig:wfstrongpert}
		\end{figure}
		Below the avoided crossing with the PEC involving the \rbminus resonance 
		($R\lesssim140$ a.u. in \figab\ref{fig:pdetail}), none of the wave functions can be characterized as 
		predominated by the $np$ states (see \figab\ref{fig:wfstrongpert}). All three of them have complicated 
		structure with significant contributions of the higher angular momenta. Correspondingly, as it is illustrated 
		in \figab\ref{fig:pdetail} for the state $12p$,  
		the local minima and maxima 
		of the oscillations in the PECs do not accurately correspond to the extremes and zeroes of the radial atomic 
		Rydberg wave functions $np$.
		
		The character of the PECs $\alpha$ and $\beta$ in \figab\ref{fig:pdetail} can be partially explained 
		by the coupling of the degenerate hydrogen-like manifolds by the neutral perturber represented by the 
		zero-range model. In that theory, the pseudopotential of the 
		perturber in the partial waves $s$ \cite{Greene-prl} and $p$ \cite{Hamilton2002,Omont1977} yields in the basis 
		of the degenerate atomic states two PECs that detach from the degenerate unperturbed asymptote.
		One of them descends with decreasing internuclear distance $R$ due to the e$^{-}$-Rb 
		resonance \cite{Bahrim2000} and becomes oscillatory at smaller internuclear separations. This corresponds to 
		the curve $\beta$ in \figab\ref{fig:pdetail}. \citet{Hamilton2002}   associated it with the butterfly 
		electronic 
		states of the LRRM. However, the probability density maps of the Rydberg electron corresponding 
		to the points of the PEC $\beta$ calculated in this work do not show the butterfly character [see 
		\figab\ref{fig:pdetail} (c) and (d)].
		
		In the energy range studied in this work, these oscillating PECs are intersected by the asymptotic energy 
		$5s+np$. This suggests that this state $np$ is coupled to the nearest degenerate manifold above it and its 
		perturbation by the neutral atom cannot be treated separately. As a result, there are no states in the studied 
		region where the perturbation of the $np$ atomic Rydberg levels would be localized to the vicinity of the 
		perturber [see \figab\ref{fig:pdetail} (a) and (b)]. This coupling is also the reason for the character of the 
		Rydberg electron probability density different from the butterfly-like shape.
		
		The authors in reference \cite{Hamilton2002} studied higher excitations of the Rydberg atoms where the 
		oscillatory segments of these deeply bound PECs are well separated from the other asymptotic thresholds. 
		Therefore, the weakly locally perturbed Rydberg states $np$ exist and the perturbed high-$l$ states show the 
		butterfly character.
		
		The role of the coupling with the states $np$ was verified by performing a set of the test one-electron 
		finite-range \rmat 
		calculations where the model potential of the perturber \cite{Khuskivadze} was artificially made more 
		attractive. The oscillatory structures in the obtained PEC $\beta$ descended towards lower 
		energies while separating from the curve $\alpha$. The PEC $\alpha$ 
		became consistent with other weakly perturbed non-degenerate states with the asymptotes $5s+n(l=0,2,3)$.
	\subsection{Trilobite-like states}
		The other type of PEC detaching from the degenerate hydrogen-like manifolds at large values of $R$ in 
		\figsab\ref{fig:lowpecs}, \ref{fig:detail11s12s} and \ref{fig:pdetail}, is monotonically increasing with the 
		nuclei approaching each other (curve $\gamma$ in \figab\ref{fig:pdetail}).

		\begin{figure}
			\includegraphics{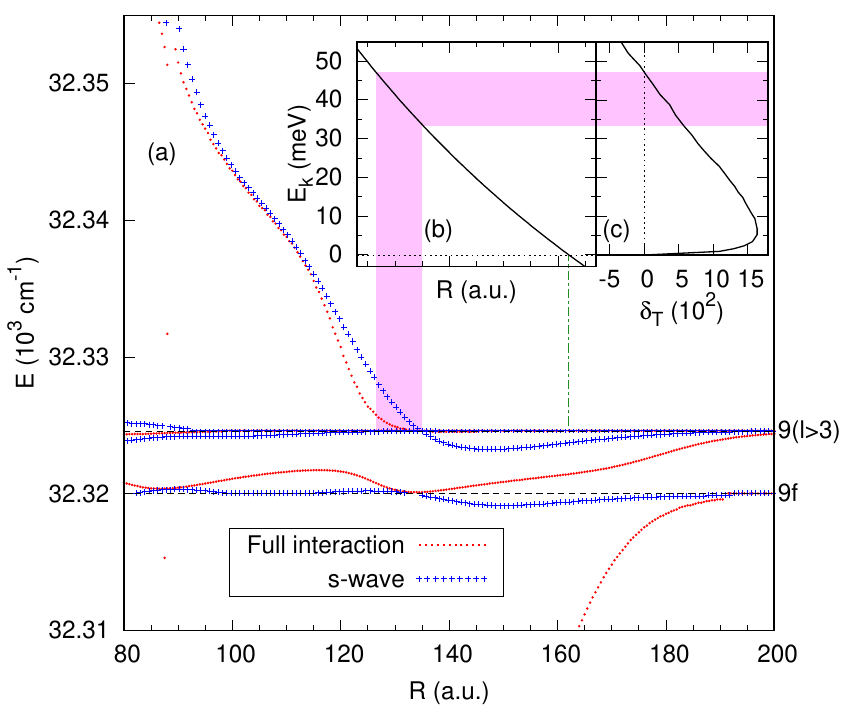}
			\caption{Detail of the PECs calculated using the finite-range one-electron model \cite{Khuskivadze} (red 
			dotted lines) near the asymptote $5s+9f$ and $5s+9(l>3)$ plotted along with the PECs obtained 
			neglecting the e$^-$-Rb interaction in the $p$-wave (blue crosses) (a). Mapping between the internuclear 
			distance $R$ and the classical kinetic energy of the Rydberg electron $E_k$ at the location of the 
			perturber (b) -- aligned by the $R$-axis to panel (a). Dependence of the $^3S^e$ phase shift on $E_k$ 
			\cite{Khuskivadze} (c) -- aligned by the axis $E_k$ to panel (b). The green dashed-dotted vertical guiding 
			line denotes the point where $E_k=0$.}
			\label{fig:pecdetail}
		\end{figure}
		The Rydberg electron probability distributions [\figab\ref{fig:wfstrongpert}(e), (f)] along this curve show the 
		dominating $s$-wave component in the vicinity of the perturber and the overall similarity to the trilobite 
		states \cite{Greene-prl, Granger}.
		
		Absence of the oscillatory structures in the PECs that support the bound 
		vibrational states is a qualitative difference from the previously studied trilobite states at higher energies 
		and larger $R$ (see \cite{Eiles2019} and references therein).
		
		In the zero-range model, the oscillating segment of the trilobite PEC spans the interval of $R$ between two 
		characteristic points. On the side of the large $R$, it is bounded by the classical turning point of the 
		Rydberg electron beyond which the perturber interacts only with the non-oscillating exponentially decreasing 
		region of the atomic Rydberg state. The left boundary of this segment
		is defined by the condition that the local kinetic energy of the Rydberg electron at the position of the 
		perturber coincides with the  Ramsauer minimum ($\approx 45$ meV) in the triplet $s$-wave e$^-$-Rb phase shift. 
		The repulsive character of the electron-atom interaction at smaller internuclear distances consequently yields 
		repulsive PEC.
		
		In order to elucidate the absence of the bound vibrational states associated 
		with the trilobite electronic wave functions emerging from the calculations presented here, an artificial
		one-electron \rmat calculation was performed where only the $s$-wave e$^-$-Rb interaction was considered. 
		The detail of the calculated PECs near the dissociation thresholds $5s+9f$ and $5s+9(l>3)$ is plotted in 
		\figab\ref{fig:pecdetail}(a) by the blue crosses. This artificial trilobite-like PEC  has a single shallow 
		minimum below the dissociation threshold $5s+9(l>3)$ at $R\approx 148$ a.u. 
		This suggests that in the range of energies and $R$ considered in this study, the attraction of the atoms due 
		to the low-energy $s$-wave e$^-$-Rb interaction is weak and it is compensated by the repulsive effect of 
		the $p$-wave interaction included in the complete models.
		
		The small extent of $R$ where the artificial trilobite-like PEC becomes attractive can also be qualitatively 
		understood in terms of the zero-range model. The atomic Rydberg
		states that are subjects of this study are relatively compact. Therefore, a small increase of the distance from 
		the positive core yields a rapid decrease of the local kinetic energy of the Rydberg electron 
		[\figab\ref{fig:pecdetail}(b)]. As a result of 
		this mapping, the e$^-$-Rb interaction in the $s$-wave is attractive 
		(and consequently induces attraction of the centers) in the narrow interval of distances from the positive core
		between the the point corresponding to the Ramsauer minimum at $127$ a.u. and the classical 
		turning 
		point at 162 a.u.
		
		As can be seen in \figab\ref{fig:pecdetail}, the point where the artificial trilobite PEC changes its nature 
		from 
		the attractive to the repulsive (at $R\approx135$ a.u.), does not accurately correspond to the Ramsauer 
		minimum in the $^3S^e$ e$^-$-Rb interaction. It is visible as the difference in $R$ and in $E_k$ marked by the 
		magenta 
		region. Since the difference of 8 a.u. is similar to the size of the ground-state Rb atom, this variation can 
		be 
		attributed to the effects of the atomic finite size. This is taken into account in the PECs calculated in this 
		work while disregarded 
		in the zero-range interpretation.
	\section{Conclusions}
		\label{sec:conc}
		The two-electron \rmat method \cite{Tarana2016} was applied to \rbtwo for a range of the internuclear 
		separations between 37 a.u. and 200 a.u. in order to calculate the excited $^3\Sigma$ electronic states of this 
		LRRM. In 
		addition to the interaction between the Rydberg electron and the neutral perturber considered also in the 
		zero-range models of the LRRMs, this approach also takes into account the effect of the Coulomb potential due 
		to the positive Rydberg core on the valence electron of the perturber. The valence electron is 
		explicitly represented and its interaction with the Rydberg electron is treated as true 
		Coulomb repulsion.
		
		The goal of the study presented in this article was to compare this advanced two-electron approach with the 
		simple zero-range model at the intermediate internuclear separations $R<200$ a.u. where these can yield 
		different energies. The PECs calculated using the two-electron \rmat method showed similar overall character to 
		those obtained from the zero-range model. The most notable differences appeared in the regions where the curves 
		detach from the asymptotic energies of the degenerate hydrogen-like Rydberg states.  Due to the low-energy 
		\electron-Rb resonance, significant probability density of the Rydberg electron is localized in the vicinity of 
		the perturber. It is not surprising that these states are sensitive to the details of all the interactions 
		involving the perturber.
		
		The two-electron \rmat technique also yields different energies than the one-electron methods in the 
		classically forbidden regions of the Rydberg electron. These differences were attributed to the fact that the 
		model \electron-Rb interaction is in the one-particle models directly or indirectly parametrized by the kinetic 
		energy of the Rydberg electron and their application in the classically forbidden region requires their 
		extension towards the negative energies. These assumptions are not required in the two-electron \rmat approach.
		
		The zero-range model \cite{Greene-prl,Hamilton2002}  and the two-electron approach also yield different results 
		at small internuclear separations among those studied in this work. This is due to 
		the polarization of the perturber by the Coulomb tail of the potential due to the positive 
		Rydberg atomic core. This effect can be in the one-electron models partially taken into account by 
		including the polarization term $-\alpha_d/2R^4$ 
		\cite{Schlagmuller2016,Schlagmuller2016-prx,Liebisch2016,Whalen2017}.
		
		The two-electron approach allowed for the calculation of the PECs associated with the excited state of the 
		neutral perturber. The curves with the asymptotic energy of the state $5p+4d$ were presented in this article. 
		The PECs can be well approximated by the atomic energy of the perturber in the state $5p$ weakly polarized by 
		the 
		distant Rydberg core. This character changes only at $R\lesssim43$ a.u.
		
		The wave functions of the Rydberg electron were calculated using the one-particle model based on the 
		finite-range potential representation of the perturber \cite{Khuskivadze}.
		
		The perturbation of the non-degenerate atomic Rydberg states $ns$, $nd$ and $nf$ (with respect to the positive 
		Rydberg center) by neutral Rb atom yields the wave functions that retain their overall 
		angular structure of the unperturbed atomic Rydberg states, the modification is localized in the vicinity of 
		the perturber. On the other hand, the global character of the perturbed atomic Rydberg states $np$ changes 
		rapidly even while the perturber is distant from the Rydberg core and the wave functions show very complicated 
		nodal structure involving high angular momenta. This is due to the fact that in \rbtwo, at the internuclear 
		separations studied in this work, the energies of the long-range molecular butterfly-like states involving high 
		angular momenta with respect to the positive core are very close to the energy levels of the atomic Rydberg 
		states $np$.

		Another category of the molecular states involving high angular momenta of the Rydberg electron, are the 
		trilobite states. In the range of the internuclear separations studied here, the PECs associated with these 
		states are anti-bonding (monotonically decreasing with the internuclear distance) and they do not possess any
		vibrationally bound states. This is due to very small interval of $R$ where the classical energy of the Rydberg 
		electron allows it to interact attractively with the rubidium perturber in the $s$-wave. Moreover, the $s$-wave 
		attraction is too weak and the overall nature is dictated by the repulsive $p$-wave.
	\begin{acknowledgments}
		The author is thankful to Roman \v{C}ur\'{i}k for stimulating discussions of this research and for the critical 
		reading of the manuscript. This work was supported by the Czech Science Foundation (Project No. P203/17-26751Y).
	\end{acknowledgments}

	\appendix*

	\section{Unphysical steeply rising PECs}
		\figsab\ref{fig:lowpecs} and \ref{fig:hipecs} show, among other, four very steeply rising PECs that do not 
		appear in the results of the zero-range model. They are artifacts of the way in which the smooth matching 
		of the inner-region and outer-region wave functions is performed on the \rmat sphere. In the case when the 
		matrix $\underline{\Gamma}$ in \eqab\eqref{eq:mmatdef} becomes singular, an additional degree of freedom 
		appears in the matching equations 
		\eqref{eq:outerbcond} and \eqref{eq:rmatdef}. The number of the matching conditions is not sufficient at those 
		singular 
		energies and the matching procedure yields an unphysical bound state.
		
		For a fixed internuclear separation $R$, these artificial bound states appear at such energies $E_x$ where 
		$\det\left[\underline{\Gamma}(E_x)\right]=0$. Note, as can be seen from \eqsab\eqref{eq:gfdef} and 
		\eqref{eq:gfpwex}, that $\underline{\Gamma}$ depends on the distance of the nuclei and on the radius $r_0$ and 
		it does not depend on the interaction inside the \rmat sphere. At each $E_x$, a vector $\mathbf{x}_0'(r_0)$ 
		with the components $(x_0')_{\overline{j}}(r_0)$ exists on 
		the sphere so that $\underline{\Gamma}(E_x)\mathbf{x}_0'(r_0)=\mathbf{0}$. As a result, 
		\eqab\eqref{eq:outerbcond} 
		yields at the energy $E_x$ for an arbitrarily selected outer-region solution $\mathbf{x}(r_0)$ more general set 
		of corresponding radial derivatives $\mathbf{x}'(r_0)$  than 
		at other energies. Specifically, when $\mathbf{x}'(r_0)$ is a vector of the radial derivatives corresponding to 
		the vector of the solutions $\mathbf{x}(r_0)$ via \eqab\eqref{eq:outerbcond}, then 
		$\mathbf{x}'(r_0)+\tau\mathbf{x}'_0(r_0)$ is also a vector of the radial derivatives corresponding to the same 
		solution $\mathbf{x}(r_0)$ for any real value of the scalar factor $\tau$.
		
		Existence of at least a single value of $\tau$ for which \eqsab\eqref{eq:rmatdef} and \eqref{eq:lincomb} are 
		satisfied, in addition to \eqab\eqref{eq:outerbcond}, is a sufficient condition for a smooth matching of the 
		outer-region solution to the inner-region wave function and, consequently, for an existence of the bound state 
		at energy $E_x$. However, this is always possible, as the additional variable $\tau$ increases the total number 
		of the variables in the homogeneous $N$-dimensional linear system formed by the \eqsab\eqref{eq:rmatdef}, 
		\eqref{eq:lincomb} and \eqref{eq:outerbcond} to $N+1$. This explains the fact, observed in the performed 
		numerical calculations, that these steeply rising PECs do not depend on the interaction inside the \rmat 
		sphere.
	
		Since these PECs are monotonic and they possess clear crossings with the other PECs, they can be easily 
		distinguished from the physical PECs and they do not represent any complication for interpretation of the 
		results. Another simple way of their identification is the identification of the energies and internuclear 
		distances where the matrix $\underline{\Gamma}$ becomes singular.
		
		These unphysical PECs descend in energy with increasing radius $r_0$. This is the reason for their absence in 
		the PECs of H$_2$ presented in TC. Since the \rmat sphere used there ($r_0=10$\thinspace a.u.) was 
		significantly smaller than used in the results presented here, all the unphysical PECs were located above the 
		studied energy interval.
%

\end{document}